\documentclass[a4paper,12pt]{article}
\usepackage{graphicx}
\usepackage{amsmath}
\usepackage{hyperref}
\usepackage{xcolor}
\usepackage{enumerate}
\title{The jump effect of a general eccentric cylinder rolling on a ramp}

\author{E. Aldo Arroyo$~\!^{(a)}$\thanks{aldo.arroyo@ufabc.edu.br} ~ and
    M. Aparicio Alcalde$~\!^{(b)}$\thanks{aparicio@ufv.br},\\
    $^{(a)}$Centro de Ci\^{e}ncias Naturais e Humanas, Universidade Federal do ABC,\\
    Santo Andr\'{e}, 09210-170 S\~{a}o Paulo, SP, Brazil.\\
    $^{(b)}$Instituto de Ci\^encias Exatas e Tecnol\'ogicas, Universidade Federal de Vi\c{c}osa,\\
    38810-000, Rio Parana\'iba, MG, Brazil.}

\date{\today}

\begin{document}
    \maketitle

    \begin{abstract}
Interesting phenomena occur when an eccentric rigid body rolls on
an inclined or horizontal plane. For example, a variety of motions
between rolling and sliding are exhibited until suddenly a jump
occurs. We provide a detailed theoretical description of the jump
effect for a general eccentric cylinder. Before the jump, when the
cylinder moves along the ramp, we can assume a pure rolling
motion. However, it turns out that when the cylinder reaches its
jumping position, both the normal and static frictional forces
approach zero. Thus, it seems that there will no longer be
sufficient force to maintain rolling without slip. In order to
have a jump without slipping, we prove that the parameters that
characterize the dynamic behavior of the cylinder must belong to
some restricted region.
    \end{abstract}

\section{Introduction}
\begin{figure}[h] \centering
    \includegraphics[width=\linewidth]{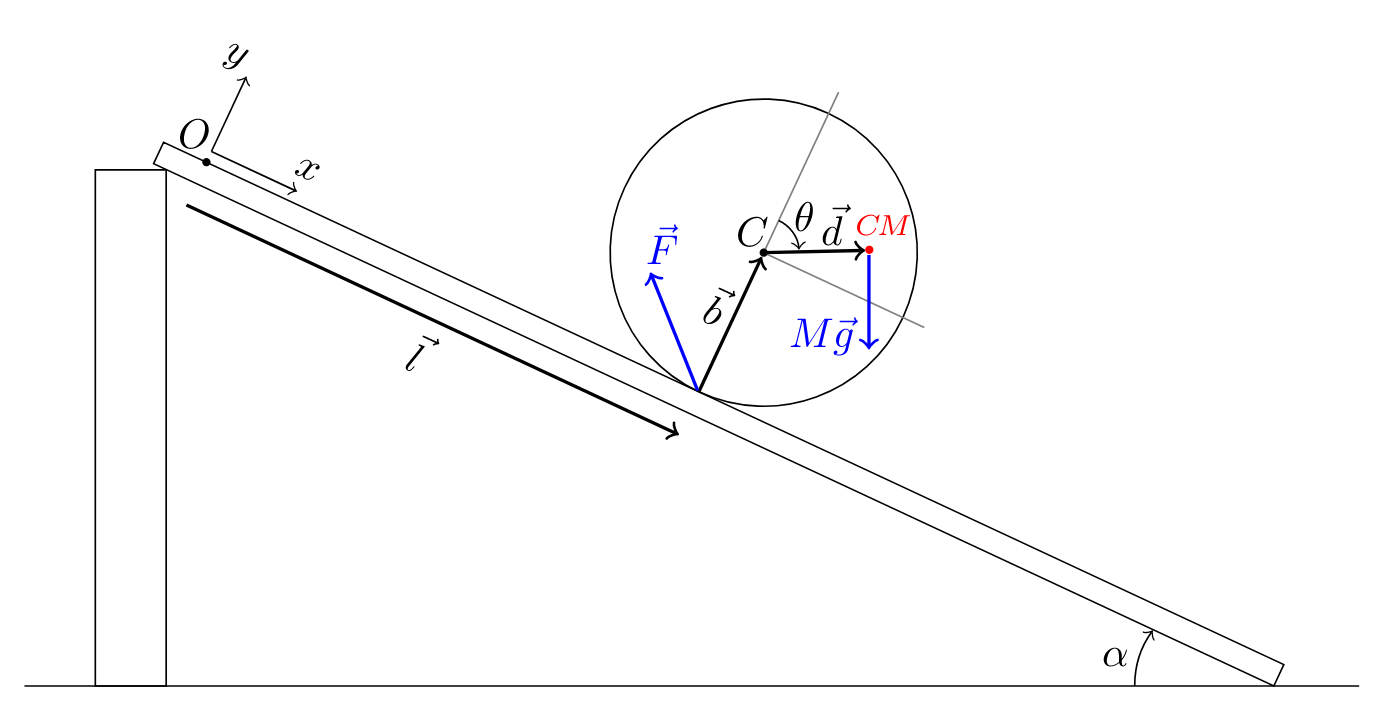}
    \caption{Schematic configuration of the physical system showing a cylinder
    rolling down a ramp of angle $\alpha$. The position of the geometric
center $C$, and the center of mass $CM$ of the cylinder with
respect to point $O$ are given by the vectors $\vec{l}+\vec{b}$
and $\vec{l}+\vec{b}+\vec{d}$, respectively. The vector $\vec{F}$
represents the force acting at the contact point between the
cylinder and the ramp, and $M\vec{g}$ is the cylinder's weight.}
    \label{fsystem1}
\end{figure}

The physical system consists of a rigid body, such as hoops,
wheels, disks, and spheres, whose center of mass is located at a
distance $d\neq 0$ from the geometric center and rolls on a
horizontal or inclined plane with friction. This system has
interesting and unexpected dynamic behavior, which has attracted
the attention of the community. In Fig. \ref{fsystem1}, a general
configuration of the physical system is exhibited, where we can
identify an inclined ramp making an angle $\alpha$ with the
horizontal, and a cylinder of radius $R$ with its center of mass
located at a distance $d$ from the geometric center. The cylinder
rolls over the ramp, and its motion is tracked by the angle
$\theta$.

In the case of a horizontal plane (i.e., $\alpha=0^\circ$), there
is literature on the dynamics of a particular cylinder. This
cylinder consists of a very thin, massless cylindrical shell with
a point mass stuck to its surface. The motion of this cylinder has
been studied and debated by several authors \cite{littlewood,
Tokieda}. When the point mass attached to the cylinder is
initially at the highest point of the cylinder, and the system is
released from rest (i.e., the initial conditions are such that
$\theta_0=0$ and $\dot{\theta}_0=0$), Tokieda \cite{Tokieda}
showed that the cylinder must jump at $\theta=90^\circ$. However,
subsequent articles have shown that the assumption made by the
author, that the jump happens immediately after the pure rolling
motion, is incorrect. For instance, it has been shown that before
the jump, there must be sliding \cite{Butler, Pritchett, yanzhu2}.
Despite this fact, Pritchett \cite{Pritchett} obtained numerical
and experimental results (for a hula hoop with a stuck point mass)
showing that the jump happens around $\theta = 90^\circ$, even
when considering slippage. Moreover, to test the jump at this
angle of $90^\circ$, Theron \cite{Theron2} claimed that the
elasticity of the hula hoop must be considered. As observed,
despite the physical system being simple, the motion of the
cylinder turns out to be, in general, complex. Further analysis
has shown that a variety of these motions can include self-induced
jumping motions \cite{TaylorAF, Bronars, Batista, Ivanov,
Shimomura}, as well as multiple transitions back and forth from
rolling to slipping \cite{Kessler, Leine, Theron, Onorato}.

In the case of an inclined plane (i.e., $\alpha\neq 0$), we have
found a few models that consider this case. One of them studied an
eccentric wheel \cite{Theron2}, and the other studied an eccentric
disk \cite{Moore2021}. These works hypothesize that the wheel (or
disk) must slip before jumping. Other references \cite{Carnevalia,
Turner, Jensen, Hu} have studied the equations of motion for this
cylinder by means of the Lagrangian formulation or a cumbersome
torque analysis. However, these works did not determine the
position where the cylinder jumps.

In all the references we have found (with the exception of
\cite{gomes}), it was assumed that the necessary condition for the
jump is when the normal force $F_y$ (acting perpendicularly from
the ramp over the cylinder) vanishes at the instant of the hop.
Let us remark that a different jump condition will be used in this
work, and this condition turns out to be equivalent to that of
\cite{gomes}. As argued in reference \cite{gomes}, the equation
that the angle $\theta$ and the angular velocity $\dot{\theta}$
must satisfy at the position where the cylinder jumps are given by
\begin{eqnarray}
    \label{jumpconditionintro}
    -g \cos\alpha + d \, \dot{\theta}^2
    \cos\theta = 0.
\end{eqnarray}
Assuming pure rolling motion, which implies conservation of
energy, and using the initial conditions $\theta_0=0$ and
$\dot{\theta}_0=0$, it was possible to find an angle $\theta_J$
that is a solution of Eq. (\ref{jumpconditionintro}).
Consequently, the position $l_J=R\theta_J$ along the inclined
plane where the cylinder jumps can be determined. However, there
is a subtlety in the pure rolling assumption. For a given value of
the coefficient of static friction $\mu_s$ between the cylinder
and inclined plane, before the cylinder reaches the position
determined by the angle $\theta_J$, it may happen that the static
friction exceeds its maximum value. This would imply that the
cylinder slips before jumping.

Therefore, aside from the jump condition issue, we remark that
there is an important open question: Are there cases where an
eccentric rigid body rolling on an inclined (or horizontal) plane
jumps just after pure rolling motion? As we are going to see, the
answer to this question is not trivial and depends on the values
of $\mu_s$, the angle $\alpha$, and the moment of inertia of the
rigid body considered. A general body that encompasses all types
of eccentric bodies having a cylindrical shape will be referred to
as a general eccentric cylinder. It turns out that the dynamic
characteristics of this cylinder can be parameterized using two
parameters, $\chi$ and $k_m$. By exploring this generality that we
consider in our work, we prove that there are (quite common)
situations in which the jump occurs immediately after sliding
roll, and there are (less common) situations where the jump occurs
immediately after pure rolling. We also prove that when the jump
occurs immediately after sliding roll, we have $F_y=0$, and in the
cases where the jump occurs immediately after pure rolling, we
have $F_y>0$. Therefore, it would be correct to consider $F_y=0$
as a necessary condition for the jump whenever there is sliding
rolling immediately before the jump, as occurs in many studies in
the literature, but it would be incorrect to state the same when
the jump occurs immediately after pure rolling or in general
cases.

This paper is organized as follows. In section 2, we will describe
the physical system to be analyzed and derive the corresponding
equations of motion. Then we will propose a model for the general
eccentric cylinder. In section 3, we determine the region where
the parameters that characterize the dynamical behavior of the
general eccentric cylinder must belong in order to have a jump
without slip, for any value of $\alpha$ and $\mu_s$. In section 4,
we will provide a summary and suggest further directions for
exploration.

\section{Description of the system and derivation of the equations of motion}
\label{system-equations}

In this section, we will define the physical system to be studied
and derive the corresponding equations of motion. The mechanical
system consists of a general eccentric cylinder rolling down an
inclined ramp, as shown in Fig. \ref{fsystem1}. The term eccentric
means that the center of mass is located at a distance of $d$ from
the geometric center. By general, we mean that the mass
distribution within the cylinder is arbitrary, with the only
restriction being that this distribution is invariant under
translations along the $z$-axis. Throughout the motion, we will
assume that the principal axis of the cylinder passing through
point $C$ always remains parallel to the $z$-axis. Note that when
the cylinder's height is very small compared to the radius, the
cylinder represents a disk, and if the mass distribution is
concentrated on the edge, the cylinder becomes a hoop or wheel. In
this sense, we can say that this general eccentric cylinder can
encompass all types of eccentric bodies that have cylindrical
shape.

In Fig. \ref{fsystem1}, we also show the coordinate system $xy$,
where the origin is set at the point $O$ located at the top of the
ramp. The $x$-axis and $y$-axis are parallel and perpendicular to
the ramp, respectively. The position of the center of mass with
respect to the geometric center of the cylinder is given by the
vector
\begin{eqnarray}
\vec{d}=d\sin\theta\,\hat{i}+d\cos\theta\,\hat{j}\,, \label{ddef}
\end{eqnarray}
where $\theta$ is the angle between the vector $\vec{d}$ and the
$y$-axis.

Using the coordinate system $xy$, which is an inertial reference
frame, we can write the position of the center of mass as follows
\begin{eqnarray}
    \vec{r}_{CM}=\vec{l}+\vec{b}+\vec{d} =
    (l+d\sin\theta)\hat{i}+(b+d\cos\theta)\hat{j}\,.
    \label{rcmdef}
\end{eqnarray}
In the case where the cylinder remains in contact with the ramp,
the vector $\vec{b}$ is given by $\vec{b}=R\,\hat{j}$, where $R$
is the radius of the cylinder. While in the case where the
cylinder loses contact with the ramp and flies, we have
$|\vec{b}|>R$.

The two equations that are used to determine the dynamics of this rigid body are:
\begin{enumerate}[(i)]
    \item Newton's second law for the motion of the center of mass:
    \begin{eqnarray}
        \vec{F}_T=M\ddot{\vec{r}}_{CM}\,,
        \label{2leidef}
    \end{eqnarray}
    where the subscript $T$ means that we are considering the total force acting on the cylinder; and

    \item Newton's second law for rotations:
    \begin{eqnarray}
        \vec{\tau}_T=I_{CM}\,\ddot{\vec{\theta}}\,,
        \label{2leitorq}
    \end{eqnarray}
    where $\ddot{\vec{\theta}}$ is defined as $\ddot{\vec{\theta}} = -
    \ddot{\theta}\,\hat{k}$, and the total torque $\vec{\tau}_T$ and moment of
    inertia $I_{CM}$ are taken around the center of mass.
\end{enumerate}
From the configuration of the system shown in Fig. \ref{fsystem1}, we can
write Eq. (\ref{2leitorq}) as follows:
\begin{align}
   -(\vec{b}+\vec{d})\times\vec{F} =
   -I_{CM}\ddot{\theta}\,\hat{k}\,.
    \label{theta-eq0}
\end{align}

Since the total force acting on the cylinder is given by
$\vec{F}_T=\vec{F}+M\vec{g}$, using Eqs. (\ref{rcmdef}) and
(\ref{2leidef}), we obtain
\begin{align}
    \vec{F}=M(\ddot{\vec{l}}+\ddot{\vec{b}}+\ddot{\vec{d}}-\vec{g})\,.
    \label{ff}
\end{align}
Let us write the force $\vec{F}$ in terms of its components in the
directions of the $x$ and $y$-axis
\begin{eqnarray}
    \vec{F}=F_{x}\hat{i}+F_y\hat{j}\,,
    \label{frampcyli}
\end{eqnarray}
note that these components $F_{x}$ and $F_{y}$ are the friction
and normal force, respectively. Now substituting Eqs.
(\ref{rcmdef}) and (\ref{frampcyli}) into Eq. (\ref{ff}), we get
\begin{align}
    &F_{x}=M\ddot{l}+Md(\sin\theta)\ddot{~}-Mg\sin\alpha\,,\nonumber\\
    &F_{y}=M\ddot{b}+Md(\cos\theta)\ddot{~}+Mg\cos\alpha\,.
    \label{ffs}
\end{align}
Using Eqs. (\ref{ddef}), (\ref{frampcyli}) and writing
$\vec{b}=b\hat{j}$, from Eq. (\ref{theta-eq0}) we obtain
\begin{align}
    I_{CM}\ddot{\theta}=d\sin\theta\,F_y-(b+d\cos\theta)F_{x}\,.
    \label{theta-eq}
\end{align}

The equations (\ref{ffs}) and (\ref{theta-eq}) will be useful in
analyzing three possible types of motion for the cylinder: pure
rolling, rolling with slipping, and flight motion. For each type
of motion, additional relations between the forces and positions
must be established. In the next two subsections, we will study
the equations in the case of pure rolling and flight motion.

\subsection{Equations in the case of pure rolling motion}
Since the cylinder is in contact with the ramp, we have $b=R$,
which means that $b$ is constant and, therefore, $\ddot{b}=0$.
Moreover, due to the pure rolling condition, we also have
$l=l_0+R(\theta-\theta_0)$, and consequently
$\ddot{l}=R\,\ddot{\theta}$. The moments of inertia $I_C$ and
$I_{CM}$ with respect to the geometric center $C$ and the center
of mass $CM$ of the cylinder are related by the equation $I_C =
I_{CM} + M d^2$. Substituting these equations into Eqs.
(\ref{ffs}) and (\ref{theta-eq}), we can derive the following
nonlinear second-order differential equation

\begin{eqnarray}
    \ddot{\theta} \left(\frac{I_{C}}{M R^2}+
    1+2 \chi \cos \theta\right)-\chi \dot{\theta}^2 \sin \theta -\frac{g}{R}  (\chi \sin (\alpha
    +\theta)+ \sin \alpha ) = 0\,,
    \label{roll-eq}
\end{eqnarray}
where we have defined the following dimensionless parameter $\chi$
as follows
\begin{eqnarray}
    \label{chi11} \chi = \frac{d}{R}\,.
\end{eqnarray}

\subsection{Equations in the case of flight motion}
In the case of flight motion, the cylinder has no contact with the
ramp, therefore the contact force is null, i.e., $F_{x}=F_y=0$.
From Eq. (\ref{theta-eq}) it is straightforward to show that the
angular velocity $\dot{\theta}$ is constant. And from Eq.
(\ref{ffs}) we obtain:
\begin{align}
    &\ddot{l}=d\,\dot{\theta}^2\sin\theta+g\sin\alpha\,,\nonumber\\
    &\ddot{b}=d\,\dot{\theta}^2\cos\theta-g\cos\alpha\,.
    \label{lb-eq2}
\end{align}
These equations are consistent with the common knowledge about the
free-fall motion of a rigid body, where the center of mass
performs a parabolic trajectory and the angular velocity is
constant.

\subsection{Transition from rolling to flight motion}
In order to understand the condition for the transition from
rolling to flight motion, we will perform the following analysis.
Before the jump, the cylinder stays in contact with the ramp, so
$b(t)=R$, $\dot{b}(t)=\ddot{b}(t)=0$, and the values of
$\theta(t)$, $\dot{\theta}(t)$, and $\ddot{\theta}(t)$ are related
by means of Eq. (\ref{theta-eq}). At the moment of the jump, the
values of $b(t)$, $\theta(t)$, $\dot{b}(t)$, and $\dot{\theta}(t)$
change continuously. The continuity of $\dot{b}(t)$ and
$\dot{\theta}(t)$ is supported by the fact that there are no
additional external forces acting on the cylinder at the instant
of the jump that would change the cylinder's momentum.

After the jump, the cylinder performs a flight motion where the
relations in Eq. (\ref{lb-eq2}) are valid. Due to the continuities
mentioned above, the values of $b=R$ and $\dot{b}=0$ are the
initial conditions for $b(t)$ in the flight motion. Therefore,
when the value of $\ddot{b}$ given by Eq. (\ref{lb-eq2}) starts to
be greater than zero, it implies that the values of $\dot{b}$ and
$b$ start to increase. It is interesting to note that this
increase in the values of $\dot{b}$ and $b$ happens because we
assumed that the cylinder is not attached to the ramp.
Consequently, at the point where the transition from rolling to
flight motion occurs, we must have $\ddot{b}=0$ in Eq.
(\ref{lb-eq2}). This condition means \footnote{This jump condition
was obtained in Ref. \cite{gomes} using an alternative approach.}:
\begin{eqnarray}
    \label{jumpcondition}
    d\,\dot{\theta}^2\cos\theta - g \cos\alpha = 0\,.
\end{eqnarray}

Before the jump, in order to understand the behavior of $\ddot{b}$
as defined in Eq. (\ref{lb-eq2}), we need to track the value of
$\dot{\theta}$. The motion of the cylinder starts with
$\dot{\theta}=0$, and due to the action of gravity, $\dot{\theta}$
increases. Therefore, at the beginning, $\ddot{b}=-g\cos\alpha<0$,
and subsequently, $\ddot{b}$ increases because of the contribution
of $\dot{\theta}^2$ (with some oscillation due to the factor
$\cos\theta$). This implies that at some future instant, when
$\ddot{b}$ approaches zero, i.e., when Eq. (\ref{jumpcondition})
is satisfied, we reach the moment when the cylinder jumps.

Let us comment that in other works \cite{yanzhu2, Theron2,
Moore2021}, the jump condition has been given by $F_y=0$. Namely,
the jump happens at the point where the normal force vanishes.
However, these works consider the hypothesis that the jump is not
possible from pure rolling, which means that the cylinder must
slide before jumping. Although the main focus of our work is not
on the slip case, using equations (\ref{ffs}) and
(\ref{theta-eq}), together with the relation $F_x=\sigma\mu_k
F_y$, where $\mu_k$ is the coefficient of kinetic friction, and
$\sigma=-1$ in case $\dot{l}>R\dot{\theta}$ (i.e., skidding
motion) and $\sigma=+1$ in case $\dot{l}<R\dot{\theta}$ (i.e.,
spinning motion), we can show that the normal force $F_y$ is given
by
\begin{eqnarray}
    \label{normalslip1}
    F_y = \frac{M \, I_{CM} \left(d \, \dot{\theta}^2 \cos \theta -g \cos \alpha \right)}{M d \left(
     \sigma \,\mu_k \sin \theta  (d \cos \theta +R)-d \sin ^2\theta\right)-I_{CM}}\,.
\end{eqnarray}

From Eq. (\ref{normalslip1}), we observe that the jump condition,
as given by Eq. (\ref{jumpcondition}), clearly implies that the
normal force $F_y$ vanishes. It should be noted that the
expression for the normal force, as given by Eq.
(\ref{normalslip1}), is only valid when there is slippage. If the
jump occurs from pure rolling motion, employing Eq.
(\ref{jumpcondition}), we will show that the normal force does not
necessarily vanish.

\subsection{Scale invariance of the dynamics and a model for the general
eccentric cylinder}

Considering a general mass distribution inside the cylinder, with
the only restriction being that this distribution is invariant
under translations along the cylinder's principal axis, such that
the center of mass does not coincide with the cylinder's geometric
center, in this subsection we analyze the equations of motion of
this general eccentric cylinder. It turns out that the equations
of motion have scale invariance. By using this invariance, we can
find common characteristics between two different cylinders in a
way that the dynamics of both cylinders are equivalent. These
common characteristics between two different cylinders can be
parameterized using two independent parameters. We also propose a
simplified model of the eccentric cylinder where the two
independent parameters have a simple geometric and mass
distribution interpretation. We prove that this simplified model
is dynamically equivalent to any general cylinder considered in
our study.

Let us consider a cartesian coordinate system fixed to the
cylinder, such that the $z$-axis coincides with the principal
axis, and the coordinate origin is located at the cylinder's
geometric center $C$. The center of mass $CM$ and the cylinder's
moment of inertia with respect to the $z$-axis are computed as
$\vec{r}_{CM}=\frac{1}{M}\int dxdydz\,\vec{r}\rho(x,y)$ and
$I_C=\int dxdydz\,(x^2+y^2)\rho(x,y)$, respectively, where
$\rho(x,y)$ is the mass density bounded by the cylindrical
surface. Since the mass distribution is invariant under
translations along the $z$-axis, the density $\rho(x,y)$ does not
depend on $z$. These specifications provide two general
implications:
\begin{enumerate}
    \item The $CM$ can be located at any distance $d$ from the cylinder's geometric center $C$,
    with this distance restricted to the interval $d\in[0,R]$. For example, the particular case
    where $d=R$ occurs when the entire cylinder's mass is distributed along a line on the
    cylindrical lateral surface that is parallel to the $z$-axis.

    \item For fixed values of $M$, $R$, and $d$, the moment of inertia $I_{CM}$
    (or equivalently $I_{C}$, thanks to the relation $I_C = I_{CM}+Md^2$),
    has minimum and maximum values, where these bounding values are:
    \begin{enumerate}
        \item {\it The minimum:} this value corresponds to the case where the whole cylinder's mass is distributed along a line
         (parallel to the $z$-axis) that passes through the $CM$, therefore in this case we have: $I_{CM}=0$ or
         $I_C=Md^2$.

        \item {\it The maximum:} this value corresponds to the case where the whole cylinder's mass is distributed
        on the cylinder's lateral surface, so in this case we have: $I_{CM}=MR^2-Md^2$ or $I_{C}=MR^2$.
    \end{enumerate}
    Given a value of $I_{C}$ such that $Md^2 \leq I_{C} \leq MR^2$, there are a variety of possibilities for the mass distribution $\rho(x,y)$ that yield the same value of
    $I_{C}$. Since for a fixed values of $M$, $R$, $d$ and $g$, the
  equations of motion, given by Eqs. (\ref{ffs}) and (\ref{theta-eq}), depend only on
  $I_{C}$, the cylinder's dynamical behavior does not depend on specific details of the mass distribution
  $\rho(x,y)$.
\end{enumerate}

From Eqs. (\ref{ffs}) and (\ref{theta-eq}), we can write the
following equation
\begin{align}
    \tilde{I}_{CM}\frac{d^2\tilde{\theta}}{d\tilde{t}^2}=\chi\sin\tilde{\theta}\,\left(\frac{d^2\tilde{b}}{d\tilde{t}^2}+
    \chi\frac{d^2\cos\tilde{\theta}}{d\tilde{t}^2}+\cos\alpha\right)-(\tilde{b}+\chi\cos\tilde{\theta})\,\left(\frac{d^2\tilde{l}}{d\tilde{t}^2}+
    \chi\frac{d^2\sin\tilde{\theta}}{d\tilde{t}^2}-\sin\alpha\right)\,,
    \label{eq-scale}
\end{align}
where we have defined the adimensional quantities
$\tilde{I}_{CM}=\frac{I_{CM}}{MR^2}$,
$\tilde{t}=\sqrt{\frac{g}{R}}t$, $\tilde{\theta}=\theta$,
$\tilde{b}=b/R$, $\tilde{l}=l/R$ and $\chi$ is given by Eq.
(\ref{chi11}). The functions $\tilde{\theta}$, $\tilde{b}$, and
$\tilde{l}$ are related to the solutions $\theta$, $b$, and $l$ of
the equations of motion. More explicitly, these relations are
given by $\theta(t)=\tilde{\theta}(\sqrt{\frac{g}{R}}t)$,
$b(t)=R\tilde{b}(\sqrt{\frac{g}{R}}t)$, and
$l(t)=R\tilde{l}(\sqrt{\frac{g}{R}}t)$. At this point, we are
ready to analyze the dependence of the cylinder's dynamics with
respect to $M$ and $R$. Consider two cylinders, one with mass and
radius given by the set $(M,R)$ and the other by $(M',R')$. If
$\chi$ and $\tilde{I}_{CM}$ are the same for both cylinders, then
due to Eq. (\ref{eq-scale}), they share the same solutions
$\tilde{\theta}$, $\tilde{b}$, and $\tilde{l}$. Therefore, the set
of functions $(\theta',b',l')$ and $(\theta,b,l)$ are related by
$\theta' = \theta$, $b' = \frac{R'}{R}b$, $l' = \frac{R'}{R}l$.
Namely, $\theta$ is equal for both sets, while $b$ and $l$ do not
depend on $M$ and scale by a factor proportional to $R$.
Similarly, since $t=\sqrt{\frac{R}{g}}\tilde{t}$, the time of the
events that happen throughout the motion (like the time when the
jump happens) scales by a factor proportional to the square root
of $R$.

From the above observations, and defining the scale factor
$\lambda=\frac{R'}{R}$, we conclude that the equations of motion
are invariant under the scale transformation $\theta' = \theta$,
$b' = \lambda\, b$, $l' = \lambda\, l$, $t'=\sqrt{\lambda}\, t$
provided that $\chi$ and $\tilde{I}_{CM}$ are the same for the two
different sets of values $(M,R)$ and $(M',R')$. Since
$\chi=\chi'$, we have that $d'=\lambda\, d$. Furthermore, from
$\tilde{I}_{CM}=\tilde{I}'_{CM}$, namely
$\frac{I_{CM}}{MR^2}=\frac{I'_{CM}}{M'R'^2}$, or
$\frac{I_{C}}{MR^2}=\frac{I'_{C}}{M'R'^2}$, and using $I_C=\int
dxdydz\,(x^2+y^2)\rho(x,y)$, we get the relation
\begin{align}
\int
dxdydz\,\left(\left(\frac{x}{R}\right)^2+\left(\frac{y}{R}\right)^2\right)\frac{\rho(x,y)}{M}=\int
dx'dy'dz'\,
\left(\left(\frac{x'}{R'}\right)^2+\left(\frac{y'}{R'}\right)^2\right)\frac{\rho'(x',y')}{M'}\,,
\nonumber
\end{align}
which means that under the transformation
$(x',y',z')=(\lambda x,\lambda y,\lambda z)$, and the change of
mass from $M$ to $M'$, we have that
$\frac{\rho(x,y)}{M}dxdydz=\frac{\rho'(x',y')}{M'}dx'dy'dz'$, or
$\frac{dm}{M}=\frac{dm'}{M'}$, that is the mass densities (divided
by the total mass) of the cylinders are related by
$\frac{\rho(x,y)}{M}= \lambda^3 \, \frac{\rho'(x',y')}{M'}$.

A summary of the previous analysis is as follows: given two
different cylinders, where to calculate their centers of mass and
moments of inertia, we use cartesian coordinate systems fixed to
them. Let $(x',y',z')$ the coordinate system fixed to one
cylinder, and $(x,y,z)$ to the other one. If these coordinate
systems are related by the scale transformation
$(x',y',z')=(\lambda x,\lambda y,\lambda z)$, and given the mass
densities $\rho(x,y)$ and $\rho'(x',y')$ such that
$\frac{\rho(x,y)}{M}= \lambda^3 \, \frac{\rho'(x',y')}{M'}$, this
condition guarantees that $\chi = \chi'$ and $\tilde{I}_{CM} =
\tilde{I}'_{CM}$, then the dynamic behavior of the cylinders are
equivalent. Since the cylinder's dynamics is basically ruled by
the two parameters $\chi$ and $\tilde{I}_{CM}$, we will propose a
particular construction of a cylinder which will be characterized
by other two parameters, where one of them can be chosen as being
the parameter $\chi$, and the second one will be related to the
parameter $\tilde{I}_{CM}$.

\begin{figure}[h]
    \centering
    \includegraphics[width=.25\linewidth]{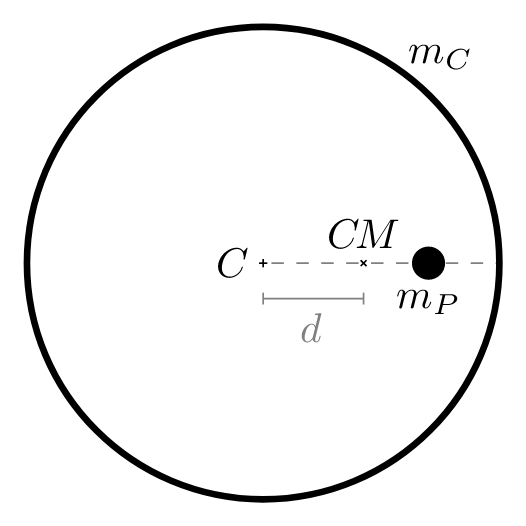}
    \caption{Cross section of a thin cylindrical shell of mass $m_C$ plus a mass line $m_P$ parallel to the cylinder's
    principal axis.}
    \label{model1}
\end{figure}

In the Fig. \ref{model1}, we show a cross section of our cylinder
model, which consists of a thin cylindrical shell of mass $m_C$
with uniform mass distribution, attached to a mass line $m_P$,
parallel to the cylinder's principal axis. Besides the theoretical
importance of this cylinder model that will be used in the
presentation of our main results, this model could also be used
for an experimental study of the jump effect.

Let us argue that this cylinder model encompasses all possible
cases of general eccentric cylinders characterized by the
parameters $\chi$ and $\tilde{I}_{CM}$. Note that the mass of the
cylinder model is given by $M=m_C+m_P$. Aside the parameter $\chi
=d/R$, where $R$ is the radius of the thin cylindrical shell, we
introduce the parameter $k_m=m_C/M$.

For example, some particular cases are:
\begin{itemize}
    \item
    One where $CM$ coincides with $C$, so we have $d=0$ (namely $\chi=0$), this case happens when the mass line $m_P$
    passes through $C$. The parameter $k_m$ controls the different possibilities for the values of the moment of inertia.

    \item
    One where $CM$ is over the border of the cylinder, it means $d=R$, this case happens when $\chi=1$ and $k_m=0$.
\end{itemize}

Through the relation that defines the $CM$, we can show that the
distance $s$ between the $CM$ and the mass line $m_P$, is given by
$s=\frac{k_m}{1-k_m}\chi R$. Then the moment of inertia $I_C$ is
computed as follows
\begin{align}
    &I_C=m_CR^2+m_P(d+s)^2=m_CR^2+\frac{M^2}{m_P}d^2\,,\nonumber\\
    &\Rightarrow \frac{I_C}{MR^2}=k_m+\frac{\chi^2}{1-k_m}\,.
    \label{imoment}
\end{align}

Since $Md^2 \leq I_{C} \leq MR^2$, we have that:
$\chi^2\leq\frac{I_C}{MR^2}\leq 1$, and therefore $0\leq\chi\leq
1$. Then from Eq. (\ref{imoment}), we have that $0\leq k_m\leq
1-\chi$. Moreover, according to Eq. (\ref{imoment}) for the
different values of $\chi$ and $k_m$ in the former intervals, we
cover all the possible values for $\frac{I_C}{MR^2}$. From Eq.
(\ref{imoment}), we have that
$\tilde{I}_{CM}=\frac{I_{CM}}{MR^2}=k_m+\frac{k_m\chi^2}{1-k_m}$,
where this parameter $\tilde{I}_{CM}$ appears in Eq.
(\ref{eq-scale}), which explicitly exhibits the independence of
the dynamics in relation to the mass $M$ and the radius $R$ when
$\chi$ and $k_m$ are fixed.

\section{Conditions for a slip-free transition from pure rolling to flight motion}

As mentioned in the previous section, the cylinder whole motion is
basically composed of three types of particular motions: pure
rolling, rolling plus slipping and flight motion. From results of
numerical simulation (work in progress \cite{progresswork}), we
have observed that depending on the initial conditions, the values
of the parameters $\chi$, $k_m$ and the ramp inclination $\alpha$,
the cylinder performs a variety of interesting motions. The most
common situation corresponds to the case where the cylinder
initially performs a pure rolling motion, then the motion is
alternated between pure rolling and rolling plus slipping, until
at some point the cylinder loses contact with the ramp and jumps.

In this section, we will study the following sequence of motions
for the cylinder and the conditions required to have such a
sequence.
\begin{align}
\theta = \theta_0 \;\; &\Rightarrow \text{initial position,} \nonumber \\
\theta_0 < \theta < \theta_J \;\; &\Rightarrow \text{pure rolling
motion,} \nonumber \\
\theta = \theta_J \;\; &\Rightarrow \text{jump position,} \nonumber \\
\theta > \theta_J \;\; &\Rightarrow \text{flight motion.}
\nonumber
\end{align}

Note that the transition from pure rolling to flight motion
happens at the point where $\theta = \theta_J$. The subscript $J$
means that we are considering the value of the quantities at the
instant of the jump. From here to the rest of the paper, we will
use the following acronym JARM to mean: jump after a pure rolling
motion.

As discussed in section \ref{system-equations}, at the point where
the cylinder jumps, the angle $\theta$ and the angular velocity
$\dot{\theta}$ must satisfy Eq. (\ref{jumpcondition}). We have
denoted by $\theta_J$ the angle that is a solution of Eq.
(\ref{jumpcondition}). Therefore from this equation, we can write
\begin{eqnarray}
    \label{jumpingeq3}
    \dot{\theta}_J^2 = \frac{g \cos \alpha }{d \cos \theta_J}.
\end{eqnarray}
Substituting Eq. (\ref{jumpingeq3}) into the equation of motion
Eq. (\ref{roll-eq}), we obtain
\begin{align}
    \label{Njump3}
    \ddot\theta_J =\frac{g (1+ \chi \cos \theta_J) \sec \theta_J \sin (\alpha +\theta_J )}{R\left(k_m+\frac{\chi^2}{1-k_m}+1+2 \chi \cos\theta_J\right)}.
\end{align}

In order to have a JARM, since by definition the cylinder does not
slip, the force of friction between the cylinder and the ramp
should be static, and
\begin{align}
    \label{ineqcondition2}
     \Big{|}\frac{F_x}{F_y} \Big{|}\leq \mu_s\,,
\end{align}
where $F_{x}$ and $F_{y}$ are the friction and normal force,
respectively. Remember that the definition of these forces as
given in Eqs. (\ref{ffs}) are valid for general rolling motions
(pure rolling or rolling plus slipping). In order to restrict our
analysis to the pure rolling motion, we should substitute the
equations $\ddot{b}=0$ and $\ddot{l}=R\,\ddot{\theta}$, into Eqs.
(\ref{ffs}); after such substitution, we obtain
\begin{align}
    \label{Njump}
    F_x &= M R \big(-\frac{g}{R} \sin\alpha -
    \chi \dot{\theta}^2 \sin\theta  + (1 + \chi \cos\theta) \ddot\theta \, \big)\,,\nonumber\\
    F_y &= M R \big(\frac{g}{R} \cos\alpha - \chi \dot{\theta}^2 \cos\theta -
    \chi \ddot\theta \sin\theta  \big)\,.
\end{align}

It is interesting to note that at the beginning of the cylinder's
motion, the kinetic energy is low (or zero if the cylinder is left
from rest), and the $CM$ of the cylinder is in a higher position
so that this initial configuration allows the cylinder to roll
down. Clearly at the beginning, $F_y$ is positive because it is
dominated by the term $Mg\cos\alpha$, subsequently the values of
$\dot{\theta}$ and $\ddot{\theta}$ grow and so the value of the
normal force approaches to zero and at some point becomes
negative. Since the cylinder is not attached to the ramp, negative
values of the normal force are not physical allowed in our study.

Before the normal force becomes negative (when it is positive and
approaching zero), the inequality given by Eq.
(\ref{ineqcondition2}) is satisfied. However, at some point, this
inequality will no longer hold, which means that the cylinder will
begin to slip. Numerical inspection \cite{progresswork} has
revealed that the point where inequality (\ref{ineqcondition2}) is
no longer valid is located in proximity to two other points: the
point where the normal force becomes zero and the point where the
jump condition is satisfied (as given by Eq.
(\ref{jumpcondition})). Based on these observations, and in order
to have a JARM in the case of pure rolling motion, we will analyze
the following condition for the normal force:
\begin{align}
\label{ineqcondition1} F_{y} > 0.
\end{align}
Since this condition is less restrictive than the condition given
by Eq. (\ref{ineqcondition2}), it is clear that we will need to
complement Eq. (\ref{ineqcondition1}) with Eq.
(\ref{ineqcondition2}).

\subsection{Conditions for a JARM independent of the initial conditions}
In order to perform a general analysis of the values and signs of
the friction and normal force at the jump point, namely when
$\theta =\theta_J$, we substitute Eqs. (\ref{jumpingeq3}) and
(\ref{Njump3}) into Eqs. (\ref{Njump}), so that we obtain
\begin{align}
    \label{jump6}
    F_{x,J} &= -\frac{g M \left(k_m+\frac{\chi^2}{1-k_m} -\chi^2 \cos^2\theta_J\right) \sec \theta_J  \sin (\alpha +\theta_J)}{k_m+\frac{\chi^2}{1-k_m}+1+2 \chi \cos\theta_J}\,,\\
    \label{Njumpfn}
     F_{y,J} &=  -\frac{g M \chi (1+ \chi \cos \theta_J) \tan \theta_J  \sin (\alpha
    +\theta_J )}{k_m+\frac{\chi^2}{1-k_m}+1+2 \chi \cos\theta_J}\,.
\end{align}

Regarding the result of the normal force $F_{y,J}$ from equation
(\ref{Njumpfn}), we note that $F_{y,J}$ is not necessarily equal
to zero. In the case where $F_{y,J} > 0$, and since for $\theta >
\theta_J$ we have $F_y = 0$, it is clear that the normal force
changes discontinuously from a non-vanishing to zero value at the
point where the cylinder jumps.

Note that some terms in Eqs. (\ref{jump6}) and (\ref{Njumpfn}) are
positive, these are: $k_m+\frac{\chi^2}{1-k_m}+1+2 \chi
\cos\theta_J>0$; and $1+ \chi \cos \theta_J>0$. Also from Eq.
(\ref{jumpingeq3}) we have: $\cos\theta_J>0$; and thus
$k_m+\frac{\chi^2}{1-k_m} -\chi^2 \cos^2\theta_J>0$. After
consideration of the positivity condition of these terms,
employing Eqs. (\ref{jump6}) and (\ref{Njumpfn}), we get
\begin{align}
    sign(F_{x,J})&=-sign(\sin(\alpha +\theta_J )),
    \label{signs1} \\
 sign(F_{y,J})&=-sign(\sin\theta_J \sin(\alpha +\theta_J )).\label{signs0}
\end{align}

A subtle issue that can be observed from equation (\ref{signs0})
is that the normal force at the jump point could even be negative.
Physically, this would mean that before the cylinder jumps, the
normal force could have been zero. Therefore, the cylinder may
have slipped before jumping, implying that the pure rolling
assumption is no longer valid. In what follows, we will address
this issue in more detail.

As mentioned before, in order to have a JARM, the condition that
the normal force should be positive needs to be complemented with
Eq. (\ref{ineqcondition2}). Therefore, let us start our analysis
by searching for possible allowed values of $\theta_J$ through Eq.
(\ref{signs0}), so that the condition given by Eq.
(\ref{ineqcondition1}) is satisfied. In that sense, it is not
difficult to show that
\begin{align}
 \label{inethetaj}
 2 \pi n -\alpha < \theta_J  < 2 \pi n, \;\;\;
 n=1,2,3,\dots
\end{align}

\begin{figure}[h] \centering
    \includegraphics[width=\linewidth]{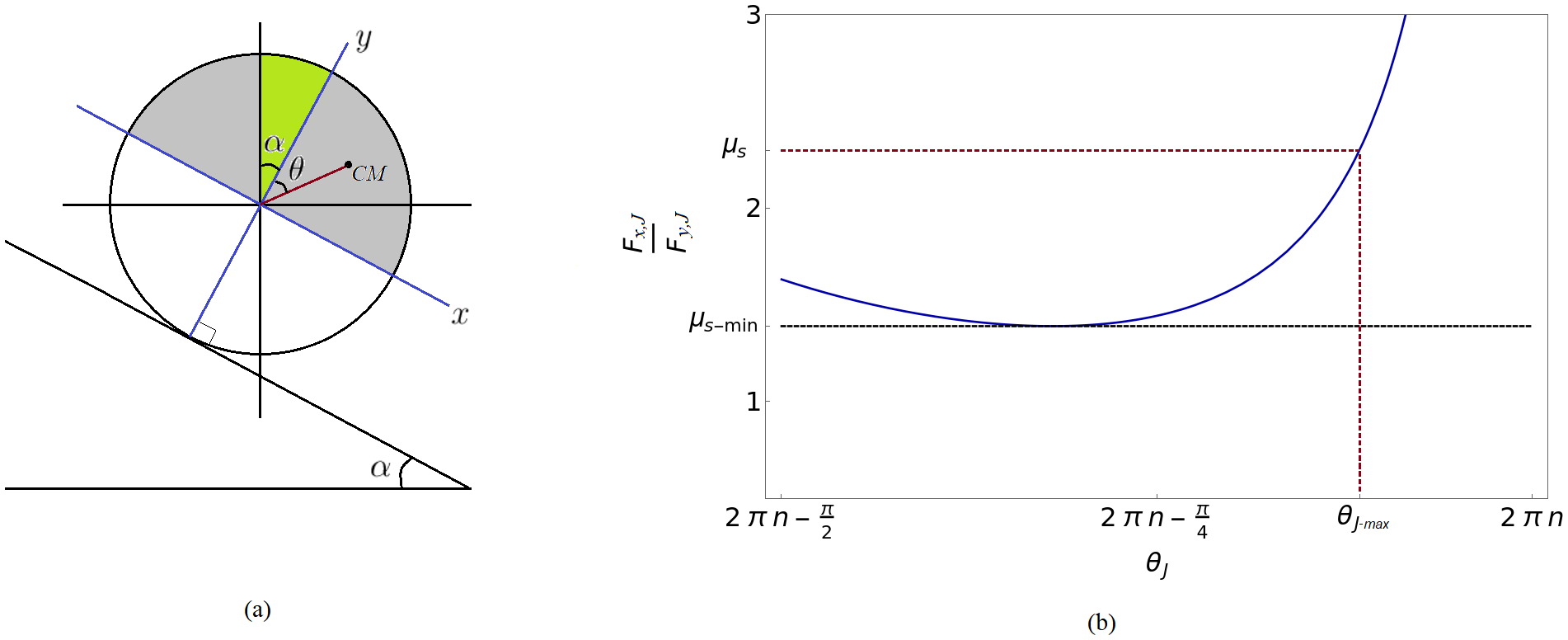}
    \caption{(a) Representation of the angles $\alpha$ and $\theta$ on the cross section of the cylinder.
    (b) Plot of the function $\mu(\chi,k_m,\theta_J)$, for $\chi=0.5$ and $k_m=0.4$.}
    \label{sign-graf}
\end{figure}

From Fig. \ref{sign-graf}a, we can see that the angle
$\alpha+\theta$ is measured between $\vec{d}$ and the vertical
line. Therefore from this geometrical configuration of the angles,
we can easily prove that the whole shadowed regions (gray and
green) corresponds to the regions where $\cos\theta>0$. Now from
Eq. (\ref{signs0}), it is not difficult to see that in the gray
region $F_{y,J}<0$, and in the green region $F_{y,J}>0$.

Therefore, the only domain where a JARM could happen (due to the
condition in Eq. (\ref{ineqcondition1})), is when the $CM$ is
inside the green region, on the other regions a JARM is precluded.
Notice that the green region agrees with the interval for
$\theta_J$ given by Eq. (\ref{inethetaj}), where the integer
number $n$ is interpreted as the counting of the full turns
completed by the cylinder. Finally, in the green region we can
check that $F_{x,J}<0$ (this result is obtained from Eq.
(\ref{signs1})), which means that the static friction force over
the cylinder points along and upward the ramp.

More restrictions for the possible allowed values of $\theta_J$
can be obtained from the condition given by Eq.
(\ref{ineqcondition2}). So, substituting Eqs. (\ref{jump6}) and
(\ref{Njumpfn}) into Eq. (\ref{ineqcondition2}) we obtain
\begin{align}
    \mu(\chi,k_m,\theta_J) \equiv \frac{\frac{-k_m^2+k_m+\chi ^2}{(k_m-1) \chi ^2}+\cos^2\theta_J}{\sin\theta_J\,\left(\frac{1}{\chi}+\cos\theta_J\right)} \leq\mu_s\,.
    \label{ineq2}
\end{align}
In the Fig. \ref{sign-graf}b, a typical plot of the function
$\mu(\chi,k_m,\theta_J)$ is shown. By fixing the value of $\mu_s$,
and using the inequality given in Eq. (\ref{ineq2}), we can
calculate the region to which $\theta_J$ belongs. It is worth
noting that the domain of $\theta_J$ as shown in Fig.
\ref{sign-graf}b is bigger than the one defined by Eq.
(\ref{inethetaj}), however it is important to keep in mind that
the lower bound for $\theta_J$ must be greater than or equal to
$2\pi n-\alpha$. We also note that due to the usual periodic
property of the trigonometric functions that appear in
$\mu(\chi,k_m,\theta_J)$, the plot of this function is the same
for any integer value $n$. After some analysis of the Fig.
\ref{sign-graf}b, we conclude that to have a JARM:
\begin{enumerate}
    \item
    There is a minimum value $\mu_{s-min}$ for the coefficient of static friction.

    \item
    For each $\mu_s>\mu_{s-min}$, we have that $max\{2 \pi n
-\alpha,\theta_{J-min}\} < \theta_J < \theta_{J-max}$; where
$\theta_{J-min}$ and $\theta_{J-max}$ are identified as follows,
when $\mu_s > \mu_{s-min}$ and $\mu_s \approx \mu_{s-min}$, there
are two solutions to the equation $ \mu(\chi,k_m,\theta_J) =
\mu_s$, these solutions are precisely $\theta_{J-min}$ and
$\theta_{J-max}$, while for values of $\mu_s$ such that $\mu_s \gg
\mu_{s-min}$, there is only one solution that is $\theta_{J-max}$.

    \item
    Since there is a maximum value for $\theta_J$, which was denoted by $\theta_{J-max}$, then there is a minimum value for
    $\alpha$, denoted by
$\alpha_{min}$ which satisfies $\alpha_{min}=2\pi
n-\theta_{J-max}$, this is true because $\theta_J$ satisfies the
inequality given in Eq. (\ref{inethetaj}).
\end{enumerate}

As a specific application of the general results presented above,
let us consider the case where $\alpha = 0$, which corresponds to
the horizontal plane. From Eq. (\ref{inethetaj}), we can see that
the only valid value for the angle $\theta_J$ in this case is
$\theta_J = 2\pi n$. However, substituting this value into the
left-hand side of Eq. (\ref{ineq2}) results in a divergence,
indicating that to maintain roll without slipping, the coefficient
of static friction $\mu_s$ must approach infinity. Since such a
value is physically impossible, we can conclude that the no-slip
condition must be violated before the cylinder jumps, regardless
of the values of $\chi$ and $k_m$. In the following discussion, we
will consider the case where $\alpha \neq 0$.

\subsection{Conditions for a JARM using initial conditions}
Our study in the previous subsection is interesting because it
proves that there are restrictions for the possible allowed values
of $\alpha$, $\mu_s$ and $\theta_J$. Also note that to derive our
previous results, we have not used any particular initial
condition.

In this subsection, we consider the following quite standard
initial conditions: the cylinder is left from rest $\dot \theta_0
= 0$ at the top of the ramp, with $\theta_0=0$. As we are going to
see, the setting of these initial conditions will further restrict
the existence of a JARM.

Using the conservation of energy, which is valid for pure rolling
motion, together with the initial conditions $\dot \theta_0 = 0$
and $\theta_0=0$, from the jump condition given by Eq.
(\ref{jumpcondition}), we can show that the angle $\theta_J$ at
the jump point is the root of the function:
\begin{align}
    \label{Jta} J(\theta) = -1 + \frac{\big(\chi  \cos \alpha -\chi  \cos (\alpha +\theta )+\theta  \sin \alpha
         \big)\cos \theta }{(2 \pi  \gamma +\cos \theta -1) \cos \alpha},
\end{align}
where the parameter $\gamma$ is defined by
\begin{align}
 \label{dega2} \gamma = \frac{k_m^2+2 k_m \chi -\chi ^2-2 \chi -1}{4 \pi  (k_m-1) \chi
 }.
\end{align}
Let us provide conditions for the existence of roots $\theta_J$ of
the function $J(\theta)$ with the restriction given by equation
(\ref{inethetaj}). It turns out that to guarantee the existence of
a root $\theta_J \in [2\pi n-\alpha, 2\pi n]$ of the function $
J(\theta)$, we must have that
\begin{align}
 \label{JtaC1} J(2\pi n-\alpha) &= \Big( \frac{\chi  \cos \alpha +(2 \pi
n-\alpha ) \sin \alpha  -\chi }{\cos \alpha +2 \pi  \gamma
-1}-1\Big) <
0, \;\;\; \text{and}, \\
 \label{JtaC2} J(2\pi n) &=\Big( \frac{n \tan \alpha }{\gamma }-1\Big) >
 0.
\end{align}

Note that through the equation: $J(\theta_J) = 0$, it should be
possible to express $\theta_J$ as a function of the parameters
$\alpha$, $\chi$, and $k_m$, namely
\begin{align}
    \label{thetasoL2} \theta_J = \theta_{n,J}(\alpha,\chi,k_m),
\end{align}
where the subscript $n$ explicitly indicates that $\theta_J$
belongs to the interval $[2\pi n-\alpha, 2\pi n]$. Using this
equation (\ref{thetasoL2}), we can write the inequality
(\ref{ineq2}) as follows
\begin{align}
  \frac{\frac{-k_m^2+k_m+\chi ^2}{(k_m-1) \chi ^2}+
  \cos^2\big(\theta_{n,J}(\alpha,\chi,k_m)\big)}{\sin\big(\theta_{n,J}(\alpha,\chi,k_m)\big) \,\Big[\frac{1}{\chi}+\cos\big(\theta_{n,J}(\alpha,\chi,k_m)\big)\Big]} \leq \mu_s\,.
    \label{ineq2L1}
\end{align}

We could think that for given values of $n$, $\alpha$, $\chi$,
$k_m$ and $\mu_s$ such that inequalities (\ref{JtaC1}),
(\ref{JtaC2}) and (\ref{ineq2L1}) are true, it would be enough to
guarantee that the cylinder will jump without first having
slipped. However, note that the above analysis has been performed
at the point where the cylinder jumps, namely the inequalities
(\ref{JtaC1}), (\ref{JtaC2}) and (\ref{ineq2L1}) do not
necessarily guarantee pure rolling motion for values of the angle
$\theta$ such that $\theta < \theta_J$. Therefore we will need to
impose more restrictions.

Since extra conditions will come from the analysis of inequalities
of the type given in (\ref{ineqcondition1}) and
(\ref{ineqcondition2}), we need to write $F_x$ and $F_y$ for
generic values of $\theta$ such that $\theta < \theta_J$. These
components of the force are given by Eqs. (\ref{Njump}).

For the initial conditions $\dot \theta_0 = 0$ and $\theta_0=0$,
using the equation of motion (\ref{roll-eq}), and the conservation
of the energy, we can express the angular velocity $\dot{\theta}$
and the angular acceleration $\ddot{\theta}$ in terms of the angle
$\theta$, so that the components of the force given in Eqs.
(\ref{Njump}) can be written as functions that depend explicitly
on $\theta$
\begin{align}
 F_x(\theta) =& -\frac{g M}{4 \chi  (2 \pi  \gamma +\cos \theta -1)^2}  \Big[  \sin \alpha \big(16 \pi ^2 \gamma ^2 \chi -16 \pi  \gamma
    \chi -4 \pi  \gamma -\chi ^2+3 \chi +2 \nonumber \\ & + 2 \pi \gamma \chi^2 +
  2 (-1 + (-3 + 6 \pi \gamma) \chi) \cos \theta + \chi \cos(
    2 \theta) - 2 \theta \sin\theta -
  4 \theta \chi \sin\theta \nonumber \\ & +
  8 \pi \gamma \theta \chi \sin \theta + \theta \chi \sin(2 \theta)\big) +\chi (2 \cos\alpha (-1 - 2 \chi +
     4 \pi \gamma \chi + \chi \cos \theta) \sin \theta \nonumber \\  \label{FxL2} & -(-2 + 4 \pi \gamma + \chi) \sin (\alpha + \theta) - \chi
((-3 + 6 \pi \gamma) \sin(\alpha + 2 \theta) +
     \sin(\alpha + 3 \theta))) \Big], \\
 F_y(\theta) =& \;\; \frac{g M}{4 (2 \pi  \gamma +\cos \theta -1)^2}  \Big[ \cos\alpha \big(6 - 16 \pi \gamma + 16 \pi^2 \gamma^2 -
   4 \chi + 2 \pi \gamma \chi \nonumber \\
 &+(-8 - 8 \pi \gamma (-2 + \chi) +
      7 \chi) \cos \theta + (2 - 4 \chi +
     6 \pi \gamma \chi) \cos(2 \theta) + \chi \cos(
    3 \theta)\big) \nonumber \\
    &-\sin \alpha (3 \theta +
   4 (-1 + 2 \pi \gamma) \theta \cos \theta + \theta \cos(
     2 \theta) - 2 \sin \theta + 4 \pi \gamma \sin \theta
     \nonumber \\  \label{FyL2}
     &+3 \chi \sin \theta + \sin(2 \theta) -
  3 \chi \sin(2 \theta) +
  6 \pi \gamma \chi \sin(2 \theta) + \chi \sin(3 \theta))\Big].
\end{align}

In order to guarantee a pure rolling motion throughout the entire
path from $\theta = 0$ to the point where the cylinder jumps
$\theta = \theta_J$, for any value of $\theta$ such that $0<\theta
< \theta_J$, the following inequalities must be satisfied
\begin{align}
    \label{extraineqcondition1}
    F_{y}(\theta) & > 0, \;\;\; \text{and}, \\ \label{extraineqcondition2} \frac{|F_{x}(\theta)|}{F_{y}(\theta)}& \leq \mu_s.
\end{align}

Let us summarize the main result of this subsection. We have shown
that using the initial conditions $\theta_0=0$ and $\dot
\theta_0=0$, to have a JARM the parameters $n$, $\alpha$, $\chi$,
$k_m$ and $\mu_s$ need to be chosen such that the inequalities
(\ref{JtaC1}), (\ref{JtaC2}), (\ref{ineq2L1}),
(\ref{extraineqcondition1}) and (\ref{extraineqcondition2}) are
satisfied. Therefore, this result imposes nontrivial restrictions
on the possible allowed values of the parameters that appear in
the equations of the problem.

\subsection{Regions in the parameter space ($\alpha$,$\chi$,$k_m$)
for fixed values of $n$ and $\mu_s$}

In order to have a JARM for the initial conditions $\theta_0=0$
and $\dot \theta_0 = 0$, in this subsection, we are going to show
the region where the parameters must belong. Since essentially we
have five parameters $n$, $\mu_s$, $\alpha$, $\chi$, and $k_m$ to
visualize the region defined by the inequalities (\ref{JtaC1}),
(\ref{JtaC2}), (\ref{ineq2L1}), (\ref{extraineqcondition1}) and
(\ref{extraineqcondition2}), we will need to fix at least two
parameters so that the remaining three parameters can be
visualized in a three-dimensional space.

\begin{figure}[h] \centering
    \includegraphics[width=\linewidth]{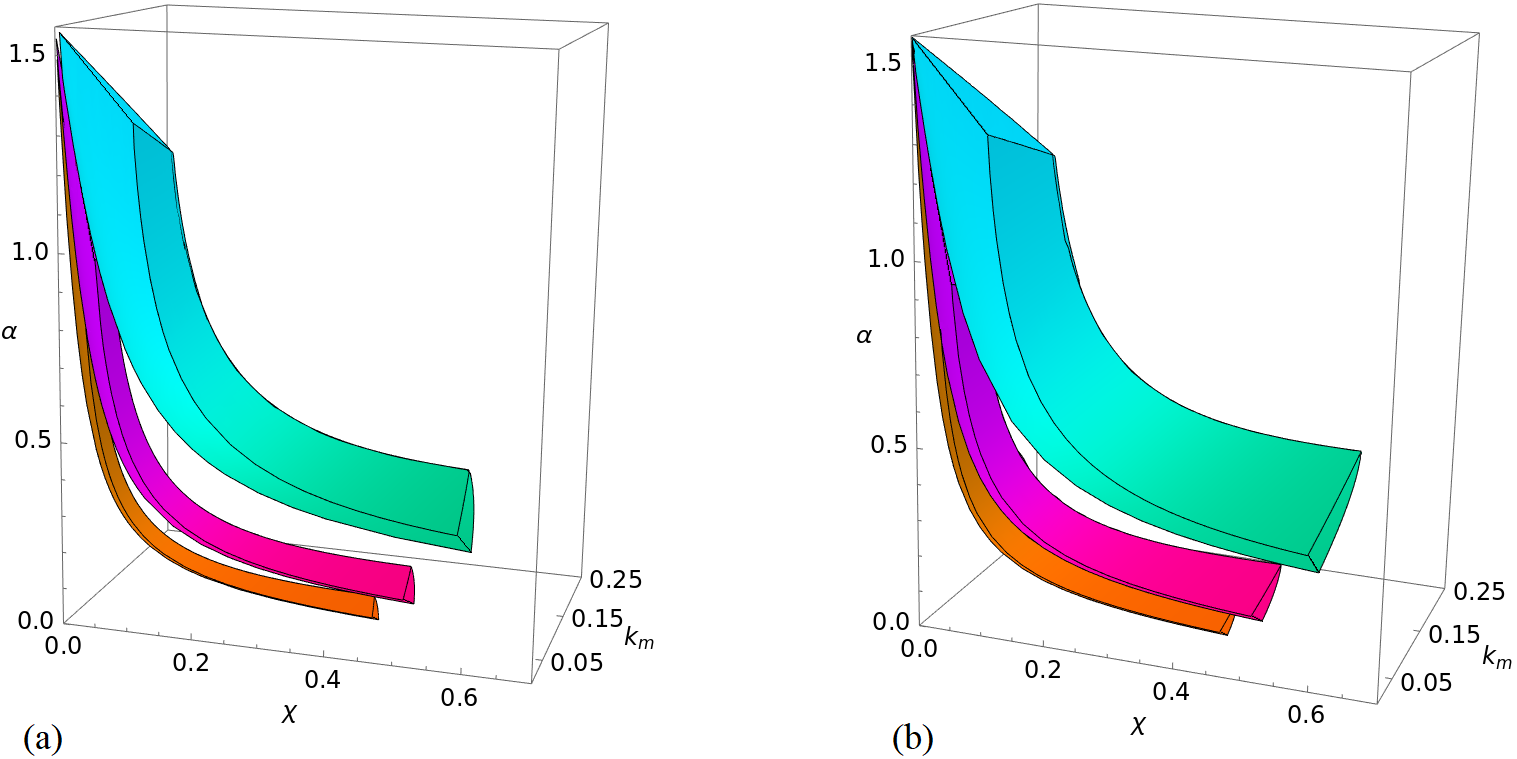}
    \caption{Regions where a JARM happens in the space
    of parameters $\alpha$, $\chi$ and $k_m$, for fixed values of
$\mu_s$, $n$ and the initial conditions $\theta_0 = 0$ and
$\dot\theta_0=0$. (a) Here we have set  $\mu_s=0.7$ and $n=1,2,3$
correspond to the green, purple and orange solids respectively.
(b) This plot is the same as the one presented in (a) but with
$\mu_s=1$.}
    \label{resultado1}
\end{figure}

By fixing the value of the parameters $n$ and $\mu_s$, in Fig.
\ref{resultado1} we present the regions in the parameter space
$(\alpha,\chi,k_m)$ where the inequalities mentioned in the
previous paragraph are fulfilled, namely if we choose any set of
parameters $\alpha$, $\chi$, and $k_m$ that belong to these
regions, we guarantee the occurrence of a JARM. Comparing Fig.
\ref{resultado1}(b) and Fig. \ref{resultado1}(a), we see that the
regions with $\mu_s=1$ are bigger than the regions with
$\mu_s=0.7$. This result makes sense since for a larger static
coefficient of friction, we expect that the cylinder has more
chances to maintain pure rolling motion.

For a fixed value of $\mu_s$, we can also compare the regions
obtained with different values of $n$. For instance, from Fig.
\ref{resultado1}(a), for the value of $\mu_s =0.7$, we observe
that as the values of $n$ increase, the corresponding regions are
getting smaller. In general, for any value of $\mu_s$, this
pattern was observed. The physical interpretation of this result
is as follows, first let us remember that $n$ represents the
number of turns performed by the cylinder, so for a given value of
$\mu_s$, since $n =1$ is the lowest possible value for $n$, the
greatest chance of having a JARM occurs before the cylinder
completes a full turn, and the chance decreases every time we
increase the value of $n$.

Regarding the inclination of the ramp given by the angle $\alpha$
(where $\pi/2$ is its maximum value), we observed that, inside the
regions shown in Fig. \ref{resultado1}, large values of $\alpha$
are in correspondence with small values of $\chi$. This type of
correspondence makes sense since for a high inclination of the
ramp, in order to avoid a slipping, the normal force should not
oscillate to much, and that happens when the value of $\chi$ is
small.

When the inclination of the ramp gets closer to the value of
$\pi/2$, we notice that the regions defined by the values of
$\chi$ and $k_m$ becomes smaller, where in the limit case $\alpha
\rightarrow \pi/2$ the parameters $\chi$ and $k_m$ vanish. This is
evidenced by the fact that when $\alpha$ is close to $\pi/2$, the
region where a JARM occurs has the shape of a wedge where its
vertex is given by the point $\alpha=\pi/2$, $\chi=0$ and $k_m=0$.
In order to interpret this result, let us remember that inside the
region where a JARM happens, large values of $\alpha$ are in
correspondence with small values of $\chi$, namely when the $CM$
is close to the geometric center $C$ of the cylinder, there are
less oscillations of the normal force, which implies in less
chances to have a slip. We conclude that high values of $\alpha$
are allowed provided that the parameters $\chi$ and $k_m$ are
small enough. Therefore, an unexpected result of having a JARM is
obtained in the limit case where $\alpha\rightarrow\pi/2$,
$\chi\rightarrow 0$ and $k_m\rightarrow 0$.

\section{Summary and discussion}
For a given value of the coefficient of static friction $\mu_s$,
we have shown that a general eccentric cylinder performs a jump
starting from pure rolling motion, provided that the angle
$\alpha$, and the parameters $\chi$ and $k_m$ that characterize
the cylinder, belong to a restricted region. If these parameters
do not belong to the aforementioned region, the cylinder has to
perform another type of motion, such as slipping with rolling,
before the jump. In a future paper \cite{progresswork}, we will
analyze these other varieties of motion using our general
cylinder.

Another important issue that can be explored is related to the
initial conditions. We have presented a general discussion about
the existence of JARM that is independent of the initial
conditions. As a result of this analysis, we show that the value
of the angle $\theta_J$ is restricted to some interval. To fix the
value of $\theta_J$, some particular initial conditions are
needed. Therefore, we have used the somewhat standard initial
conditions, $\theta_0=0$ and $\dot \theta_0 =0$. It will be an
interesting and non-trivial problem to analyze how the regions
shown in Fig. \ref{resultado1} change when other initial
conditions are used. For example, we can see if these regions
increase or decrease, which would physically imply a greater or
lesser chance of having JARM. What would be the optimal initial
conditions that allow a greater chance of having JARM?

Finally, since only the slipping motion before the jump has been
observed in experiments and theoretical studies carried out to
date, it has been a common conclusion that the no-slip conditions
must be violated before the jump. Indeed, when $\alpha=0$, on
general grounds, we have definitely shown that this last
conclusion is true. However, in the case where $\alpha \neq 0$,
there is a chance to have JARM. Therefore, to empirically test the
occurrence of JARM, it would be important to set up an experiment
that takes into account appropriate values for $\mu_s$ and the
parameters $\chi$ and $k_m$.

\section*{Acknowledgements}
We would like to thank Dominique Sugny, Gabriela and Pavao
Marde\ifmmode \check{s}\else \v{s}\fi{}i\ifmmode \acute{c}\else
\'{c}\fi{} for useful discussions.



\bibliographystyle{ieeetr} 
\bibliography{bibcyli-1} 
\end{document}